\documentclass[11pt,,showpacs]{article}  

\usepackage{amssymb,amstext,amsmath,amsthm}
\usepackage[dvips]{graphicx}
\usepackage{latexsym}
\usepackage{psfrag}
\usepackage{amsfonts}
\usepackage{bbm}

\newcommand{\Ref}[1]{(\ref{#1})}

\newcommand{\be}{\begin{equation}}
\newcommand{\ee}{\end{equation}}
\newcommand{\barray}{\begin{array}}
\newcommand{\earray}{\end{array}}
\newcommand{\bea}{\begin{eqnarray}}
\newcommand{\eea}{\end{eqnarray}}
\newcommand{\bs}{\begin{subequations}}
\newcommand{\es}{\end{subequations}}
\newcommand{\balign}{\begin{align}}
\newcommand{\ealign}{\end{align}}
\newcommand{\equ}{\begin{equation}}
\newcommand{\nequ}{\end{equation}}
\newcommand{\eqa}{\begin{eqnarray}}
\newcommand{\neqa}{\end{eqnarray}}
\def\nn{\nonumber}
\newcommand{\vj}{\vec{\jmath} }

\newcommand{\bra}[1]{\la {#1}|}
\newcommand{\ket}[1]{|{#1}\ra}

\newcommand{\id}{\mathbbm{1}}

\def\la{\langle}
\def\ra{\rangle}

\makeatletter
\newcommand*{\simboloG}[1]{%
  \vphantom{\sum}
  \smash{%
    \mathchoice{%
      \raisebox{-.3\height}{\Huge$\m@th\displaystyle#1$}%
      }
      {%
      \raisebox{-.1\height}{\Large$\m@th#1$}%
      }{%
      \raisebox{-.1\height}{\small$\m@th#1$}%
      }{%
      \raisebox{-.1\height}{\LARGE$\m@th#1$}%
      }%
    }}
\newcommand{\BigTimes}{\mathop{\simboloG{\times}}}
\makeatother
\makeatletter

\newcommand*{\simboloB}[1]{%
  \vphantom{\sum}
  \smash{%
    \mathchoice{%
      \raisebox{-.1\height}{\Large$\m@th\displaystyle#1$}%
      }
      {%
      \raisebox{-.1\height}{\Large$\m@th#1$}%
      }{%
      \raisebox{-.1\height}{\small$\m@th#1$}%
      }{%
      \raisebox{-.1\height}{\LARGE$\m@th#1$}%
      }%
    }}
\newcommand{\bigTimes}{\mathop{\simboloB{\times}}}
\makeatother

\newcommand{\Z}{\mathbb{Z}}

\newcommand{\R}{\mathbb{R}}
\DeclareMathOperator{\tr}{Tr}

\def\to{\rightarrow}

\def\d{\delta}
\def\f{\frac}
\def\tl{\tilde}
\def\tn{\tilde{n}}
\def\tz{\tilde{z}}

\def\tz{\tilde{z}}
\def\tX{\tilde{X}}

\def\si{\sigma}

\usepackage{bbm}
\def\Id{{\mathbbm 1}}
\def\C{{\mathbbm C}}
\def\R{{\mathbbm R}}
\def\Z{{\mathbbm Z}}

\newcommand{\SU}{\mathrm{SU}}

\newcommand{\su}{\mathfrak{su}}

\let\eps=\epsilon
\def\th{\theta}
\def\al{\alpha}

\newcommand{\tN}{\tl{N}}

\newcommand{\bz}{\bar{z}}

\begin{document}

\title{ From twistors to twisted geometries}
\author{{Laurent Freidel${}^{a}$ and Simone Speziale${}^{b}$}\footnote{lfreidel@perimeterinstitute.ca, simone.speziale@cpt.univ-mrs.fr}
\smallskip \\ 
{\small ${}^a$\emph{Perimeter Institute for Theoretical Physics, 31 Caroline St. N, ON N2L 2Y5, Waterloo,Canada}} \\ 
{\small ${}^b$\emph{Centre de Physique Th\'eorique,\footnote{Unit\'e Mixte de Recherche (UMR 6207) du CNRS et des Universites Aix-Marseille I, Aix-Marseille II et du Sud Toulon-Var. Laboratoire affili\'e \`a la FRUMAM (FR 2291).} CNRS-Luminy Case 907, 13288 Marseille Cedex 09, France}}}

\maketitle

\begin{abstract}
In a previous paper we showed that the phase space of loop quantum gravity on a fixed graph can be parametrized in terms of twisted geometries, quantities describing the intrinsic and extrinsic discrete geometry of a cellular decomposition dual to the graph. Here we unravel the origin of the phase space from a geometric interpretation of twistors.
\end{abstract}

\section{Introduction}

The phase space of loop gravity on a fixed graph is given by holonomies of the gravitational connection and fluxes of the triad field. In \cite{noi}, we introduced a parametrization of this phase space in terms of quantities describing the intrinsic and extrinsic discrete geometry of a cellular decomposition dual to the graph. The description provides a natural extension of Regge geometries allowing for discontinuous metrics \cite{noi} (See also \cite{IoCarlo,BiancaJimmy}).
The name \emph{twisted} was meant to stress this discontinuous nature, but also to imply the existence of a relation to twistors. In fact, as  we show explicitly in this brief note, the parametrization can be derived from a geometric interpretation of twistors.

\medskip

\section{Twisted geometries from phase space reduction}\label{SecRed2}

Our starting point is the twistor space 
\be
 \mathbb{T} \equiv  \C^{2} \times \C^{2},
\ee
with coordinates $(z_A,\tz_A)$, $A=0,1$. We equip $ \mathbb{T}$ with the standard Poisson algebra,
\be\label{Pz}
 \{z_{A},\bar{z}_{B}\} = - i  \d_{AB},\qquad \{\tilde{z}_{A},\bar{\tilde{z}}_{B}\} = - i \d_{AB}.
 \ee
In each $\C^2$ space we introduce the 2-dimensional spinors $|{\bf{z}}\ra\equiv (z_{0},z_{1})$ and $|{\bf{z}}]= (-\bar{z}_{1},\bar{z}_{0})$.
Both spinors can be used to construct a 4-dimensional future-pointing null vector $X^{\mu}=(X^0,X^i)$. Choosing the first one, we have
\be\label{3}
|{\bf z} \ra\la {\bf z}| =  X^{0}\id +  X^{i} \sigma_{i},
\ee 
where $\si_i$ are the Pauli matrices.
In components, 
\be\label{Vz}
X^{0} = \frac12(|z_{0}|^2 + |z_{1}|^2)\equiv \frac12 \la {\bf z}|{\bf z} \ra,\quad
X^{+} = \bar{z}_{0} z_{1},\quad 
X^{-} = z_{0} \bar{z}_{1},\quad
X^{3} = \frac12(|z_{0}|^2 - |z_{1}|^2),
\ee
with
 $X^i\equiv \tr(X\sigma^i)$, and\footnote{Our conventions imply that 
  $\sigma_{3}=\sigma^{3}$, $\sigma_{-}=\sigma^{+}/2$, $\sigma_{+}=\sigma^{-}/2$,
so that  the scalar product in these components reads $X^3 Y^3 + X^+ Y^-/2 + X^- Y^+/2 $.
}
 $\sigma^{\pm} =\sigma_{1}\pm i\sigma_{2}$. Notice that
\Ref{Vz} is nothing but the classical version of the well-known Schwinger representation of the angular momentum in terms of two harmonic oscillators.
We can then parametrize $\C^2_*=\C^2\backslash \{\la {\bf z}| {\bf z}\ra = 0\}$ in terms of the null vector $X^\mu$ and a phase,  
$\varphi~\equiv~\arg(z_0) + \arg (z_1)$ 
(which is well defined provided $\la {\bf z}| {\bf z}\ra \neq 0$),
\be
\C_{*}^{2}=\{(X^i, \varphi) \}.
\ee
The induced algebra reads
\begin{subequations}\label{PPT1}\be\label{PP1b}
\{X^i,X^j \} = \eps^{ij}{}_k X^{k},
\ee
\be
\{X^{0},\varphi\}=1, \qquad \{ X^3,\varphi \} = 0, \qquad \{ X^\pm,\varphi \} = \f{X^0}{X^\mp}.
\ee\end{subequations}
Similarly, we denote $\tX^\mu$ the null vector built from $\tz_A$ as $|{\bf\tz}\ra\la {\bf\tz}| = \tX^{0}\id +\tX^{i}\sigma_{i}$, and $\tl\varphi$ the left over phase.
This leads to parametrize $\mathbb{T}_{*} = \C^2_*\times\C^2_*$ as
\be\label{T1}
 \mathbb{T}_{*} = \{ (X^i, \tX^i, \varphi, \tl \varphi) \},
\ee
where both $(X^{i},\varphi)$ and $(\tX^{i},\tilde{\varphi})$ satisfy the same algebra \Ref{PPT1}, while they commute with each other.

Consider now the constraint
\be\label{H}
H\equiv X_{0}-\tX_{0}=0,
\ee
imposing the two spatial vectors to have the same norm. This constraint generates the following U(1) action on $\mathbb{T}$,
\be\label{U(1)}
\{H, {z_A} \} = \f{i}2z_A, \qquad \{H, {\tz_A} \} = -\f{i}2 \tz_A, \qquad 
(\ket{\bf z}, \ket{\bf {\tl z}}) \mapsto (e^{i\f\th2}\ket{\bf z}, e^{-i\f\th2}\ket{\bf {\tl z}}),
\ee
which  leaves $X^{i}$ and $\tX^{i}$ invariant, while it translates the angles, 
\be
\varphi \to \varphi+\theta,\qquad \tilde{\varphi} \to \tilde{\varphi}- {\theta}.\ee
We claim that the symplectic reduction of the eight-dimensional twistor space $\mathbb{T}_*$ by the constraint \Ref{H} gives the six-dimensional phase space of twisted geometries
\be\label{defP}
P_* \equiv  S^2_j\bigTimes S^2_j \bigTimes T^*S^1 \, = \, \{(N, \tN, j,\xi)\}\backslash \{j=0\},
\ee
where $N$ and $\tN$ are unit vectors parametrizing the two spheres of radius $j\in\R\backslash \{0\}$, and $\xi$ is an angle.

Let us make this statement more precise. Recall \cite{noi} that \Ref{defP} is a symplectic space locally isomorphic to the cotangent bundle of SU(2),\footnote{We parametrize $T^*\SU(2)\cong \su(2)\times \SU(2)$ with a pair $(X,g)$. The isomorphism can be made global, i.e. including the configurations $j=0$ and $|X|=0$, taking an appropriate closure of $P_*$, see \cite{noi} for details.}
\be\label{TSU2P}
P_*/\Z_2 \cong T^*\SU(2)\backslash \{|X|=0\},
\ee
where the quotient by $\Z_2$ corresponds to the identification 
\equ\label{Z2}
( N,\tilde{N}, j, \xi)\leftrightarrow (-N,-\tilde{N},-j, -\xi).
\ee
We can now make the following

{\bf Proposition 1:} 
\be\label{TP}
\mathbb{T}_{*}/\!/{\rm U}(1) \cong P_*.
\ee 

{\bf Proof.} To prove it, it suffices to consider one of the two branches $j\gtrless 0$ identified by \Ref{Z2}. We consider $j>0$, but the proof is analogous for $j<0$. 
Let us denote by $j>0$ the common norm of the vectors $X^{i}$ and  $\tilde{X}^{j}$,
\begin{subequations}\label{isom}\be\label{defj}
 j\equiv\frac12(X^{0}+\tilde{X}^{0}), 
\ee
and introduce the unit vectors 
\be\label{defNN}
N^i = \f{X^i}{j}, \qquad \tN^i = \f{\tX^i}{j}.
\ee
In order to make contact between the original variables $z_A$ and \Ref{defNN}, we need to 
 parametrize the vectors on the sphere as $N(z)$ in terms of the stereographic complex coordinate $z$. For instance using the conventions of \cite{noi},
 $$N^i(z)= \frac1{(1+|z|^{2})}\Big( (1-|z|^{2}) , -2z, -2\bar{z}\Big), \qquad i=(3,-,+)$$
 and the same for $\tN(\tz)$. Then taking \Ref{Vz}, we see that \Ref{defNN} is achieved through the Hopf maps $z\equiv -\bar{z}_1/\bar{z}_0$, $\tz\equiv-\bar{\tz}_1/\bar{\tz}_0$.
 
The variables $j$, $N^i$ and $\tN^i$ span a 5-dimensional subspace commuting with the constraint \Ref{H}. 
Hence, it only remains to identify the sixth and last variable spanning the reduced phase space. To do so, we evaluate
$$
\{i \ln\f{z_A}{\bar z_A}, H \} = 1, \qquad \{i\ln\f{\tz_A}{\bar \tz_A}, H \} = -1, \qquad
\{i\ln\f{z_A}{\bar z_A}, j \} = \f12, \qquad \{i\ln\f{\tz_A}{\bar \tz_A},j \} = \f12.
$$
From these brackets it follows that if we define
\be\label{isom3}
\xi_A \equiv i \left(\ln\f{z_A}{\bar z_A} + \ln\f{\tz_A}{\bar \tz_A}\right),
\ee\end{subequations}
we have
\be\label{PP1a}
\{\xi_A, H\} = 0, \qquad \{\xi_A, j\} =1.
\ee
That is, both $\xi_0$ and $\xi_1$ commute with the constraint, and furthermore are conjugated to $j$. 
They are thus equally valid choices for the reduced space, 
related by the canonical transformation \linebreak $\xi_{1}=\xi_0 + 2\arg(z)+2\arg(\tz)$. 

  We conclude that the reduced phase space is spanned by $(N(z), \tN(\tz), j, \xi_A)$.   
Concerning its Poisson algebra, we have the right bracket of \Ref{PP1a}, as well as the brackets \Ref{PP1b} written in terms of \Ref{defNN}. It is also immediate to see that $j$ commutes with both $N$ and $\tN$. The only remaining brackets to evaluate are
\be\label{PPL}
\{ \xi_A, jN^i\} \equiv  L^i_A(N),
\ee
which give, in cylindrical components $(i=3,-,+)$, 
\be
L^i_0\big(N(z)\big)=(1,-z, -\bar{z}), 
\qquad L^i_1\big(N(z) \big) = (1,1/\bar{z},1/{z}).
\ee
Here $L(N) \equiv L_0( N(z)) $ is precisely the Lagrangian introduced in \cite{noi}, 
and $L_{1}(N)=L(N(-1/\bar{z}))=L(-N(z))$.
From now on, we take $\xi\equiv\xi_0$ as the reduced variable. As explained in \cite{noi}, the existence of canonical transformations which shift the $\xi$ variable and the Lagrangian are related to changes of section in the Hopf map.

Collecting the brackets \Ref{PP1b}, \Ref{PP1a} and \Ref{PPL}, we find
\begin{subequations}\label{PP}\eqa\label{PS2}
&& \{jN^i, jN^j \} = \eps^{ij}{}_k\,j N^k,
\hspace{.8cm} \{j \tl N^i, j \tl N^j \} = \eps^{ij}{}_k\, j \tl N^k, 
\hspace{.7cm} \{N^i, \tl N^j \} = 0, \\ \nn\\\label{Pzero}
&& \{\xi, j\} = 1, \hspace{2.95cm} \{N^i, j \} = 0, \hspace{2.6cm} \{\tl N^i, j \} = 0, \\ \nn\\\label{PL}
&& \{\xi, j N^i \} \equiv  L^i(N), \hspace{1.65cm} \{\xi, j \tl N^i \} \equiv  L^i(\tl N),\label{PTh}
\neqa\end{subequations}
which can be recognized as the algebra of twisted geometries, with both spheres positively oriented.\footnote{With respect to the opposite orientation taken in \cite{noi}, this different choice affects the isomorphism with $T^*\SU(2)$ in a minor way, see \Ref{mapg} below.}
$\square$

\medskip
The proof shows how the algebra \Ref{PP} descends in a simple way from the canonical Poisson brackets on twistor space. 
We remark also that in the spirit of the Guillemin-Sternberg theorem \cite{GS}, the symplectic quotient $P_{*} $ can be written as a complex quotient \emph{without} imposing the constraints:
\be
\mathbb{T}_{*}/\!/{\rm U}(1) \cong P_{*}\cong  \mathbb{T}_{*}/\C,
\ee
where the $\C$ action is given by:
$$(\ket{\bf z}, \ket{\bf {\tl z}}) \mapsto (\lambda \ket{\bf z}, \lambda^{-1}\ket{\bf {\tl z}}).$$
It is indeed trivial to show that we can always reach the constraint surface by choosing \linebreak
$\lambda =  \sqrt{\la {\bf \tz}|{\bf \tz} \ra}/\sqrt{\la {\bf z}|{\bf z}\ra}$.

Let us go back to the symplectomorphism \Ref{TSU2P}, and notice that together with \Ref{TP}, it implies the symplectic reduction from twistor space to the cotangent bundle of SU(2). For completeness, we now give explicitly this alternative reduction. 

{\bf Proposition 2:}
\be\label{TTSU2}
\mathbb{T}_{*}/\!/U(1) \cong T^*\SU(2)\backslash\{|X|=0\}.
\ee

{\bf Proof.} Recall \cite{noi} that if we trivialize $T^*\SU(2)\cong \su(2)\times \SU(2)$ as $(X,g)$ with right-invariant vector fields $X$, we have that 
\be\label{deftX}
\tX \equiv - g^{-1} X g
\ee
is a left-invariant vector field and that the Poisson algebra on linear functions reads
\begin{align}\label{PT*G1}
& \{X^i, X^j \} = \eps^{ij}{}_k X^k, & \{\tX^i, \tX^j \} = \eps^{ij}{}_k \tX^k, 
&& \{X^i, g \} = -\tau^i g, && \{\tilde X^i, g \} =  g \tau^i.
\end{align}
The first two brackets hold automatically in the reduction of $\mathbbm{T}_*$, since $X^i$ and $\tX^i$ commute with $H$ and satisfy \Ref{PP1b}.
It thus suffices to find $g(z_A,\tz_A)$ in $\mathbbm{T}_*$ such that $(i)$ it is an SU(2) group element, $(ii)$ it commutes with $H$, and ($iii$) it satisfies \Ref{deftX} and \Ref{PT*G1}. It is not hard to see that
\be\label{gzz}
g(z_A,\tz_A) \equiv \frac{\ket{\bf z}[{\bf \tz}| - |{\bf z}]\bra{\bf \tz}}{ \sqrt{\la {\bf z}|{\bf z} \ra\la {\bf \tz}|{\bf \tz} \ra}},
\ee
fulfills $(i-iii)$. Indeed, thanks to $\la{\bf \bz}|{\bf z} ]=0$, one can check that this map satisfies 
\be
g|{\bf{\tz}}] = |{\bf{z}} \ra,\qquad g |{\bf{\tz}}\ra = - |{\bf{z}} ],\qquad  gg^{\dag }= g^{\dag} g =\Id.
\ee
The commutation with $H$ is straightforward. A less trivial calculation shows also that the  matrix elements commute among themselves when $H=0$ is satisfied.
 Finally, \Ref{deftX} follows from 
 $g{|\bf{\tz} ][\bf{\tz}} |g^{\dag} = {\bf |z\ra\la z|}$, and
\Ref{PT*G1} from the brackets \Ref{Pz} and the parametrization \Ref{Vz}. $\square$

\section{Null twistors}

Thus far, we have connected the twisted geometries to pairs of spinors in $\C^4$. We now show that our construction is effectively related to \emph{twistors}, in particular to null twistors. To that end, let us briefly review some basic facts about twistors, refering the reader to the literature \cite{PR} for details.
A twistor $Z^{\alpha} \in \C^{4}$ can be viewed as a pair of spinors $Z^{\alpha}=(|\omega \ra, |\pi\ra)$, where
$|\pi \ra $ defines a null direction $ p_{\pi}= |\pi ][\pi|$ in Minkowski space, while $ |\omega \ra $ defines a 
point $x$ in complexified Minkowski space 
via $|\omega \ra = ix |\pi\ra$. 
On twistor space there is a natural hermitian pairing given by
$$\bar{Z}_{\alpha} Z^{\alpha}= \la \omega |\pi\ra + \la \pi |\omega \ra,$$
and the quantity $s=\bar{Z}_{\alpha} Z^{\alpha}/2$ is called the helicity of the twistor.
When a twistor is null, i.e. $s=0$, the matrix $x$ is Hermitian and thus identifies a point in real Minkowski space.
However, $x$ is defined only up to the addition of a null momentum $p_{\pi}$, since $p_{\pi}|\pi\ra =0$. 
The resulting null ray $x+\lambda p_\pi$ can be explicitly reconstructed as
\be
x(\lambda)= \frac{|\omega\ra \la\omega |}{i \la \omega| \pi \ra}  + \lambda |\pi][\pi|, \qquad \lambda \in \R.
\ee
Hence, a null twistor defines a null generator $p_\pi$ and a null ray in Minkowski space.
We call these data a ``ruled'' null ray, since the ray has a specific generator.

The relation between twistors and twisted geometries is established through the map
\be
|\omega \ra \equiv  |{\bf z}\ra + |{\bf \tz} ], \quad |\pi \ra \equiv  |{\bf z}\ra - |{\bf \tz} ].
\ee
Under this map the twistor Hermitian pairing becomes 
\be\label{s}
s = \frac12 \Big(\la \omega | \pi \ra + \la \pi |\omega\ra \Big)  = \la {\bf z} | {\bf z} \ra - [{\bf \tz} | {\bf \tz} ].
\ee
Then, the constraint $H=0$ in \Ref{H} is equivalent to the statement that $Z^{\alpha}({\bf z},{\bf \tz})$ is a null twistor, and
the U(1) action \Ref{U(1)} translates into a global rescaling of $Z^{\alpha}$:
\be\label{Twiphase}
Z^{\alpha}=(|\omega \ra, |\pi\ra) \to (e^{i\frac{\theta}2}|\omega \ra, e^{i\frac{\theta}2} |\pi\ra)= e^{i\frac{\theta}2}Z^{\alpha}.
\ee
Therefore $P_*$, which is the symplectic reduction of the space $\{(\ket{\bf z},\ket{\bf \tz})\}$ by $H=0$, can be interpreted as a
phase space of null twistors $\mathbb{T} \mathbb{N}$ up to a global phase,
\be\label{PTN}
P_* = \mathbb{T} \mathbb{N}/ \mathrm{U}(1).
\ee

This is the connection between (null) twistors and twisted geometries. Notice that the U(1) rescaling \Ref{Twiphase} leaves invariant the ruled null ray $x+\lambda p_\pi$ defined by $Z^\al$, thus \Ref{PTN} means that an element of $P_*$ defines a ruled null ray.
The reverse is also true: Given a null ray in Minkowski space with a specific null generator,  we can reconstruct 
uniquely a null twistor up to a global phase, and hence an element of the phase space $P_*$.

This mathematical correspondence shows that we can think of an element of $P_*$, the edge phase space of loop quantum gravity, as a ruled null ray in Minkowski space.
Whether this is just a mathematical correspondence, or it has a deeper geometrical origin, is still a mystery for us, and a fascinating one.

\section{Geometrical meaning of the constraints}

To understand the geometrical meaning of the constraints $H_e$, consider a cellular decomposition dual to the graph. A twisted geometry assigns to each face (dual to the edge $e$)
its oriented area $j_e$, the two unit normals $N_e$ and $\tN_e$ as seen from the two vertex frames sharing it, and an additional angle $\xi_e$ related to the extrinsic curvature between the frames. 
Working with $\C^4{}_e=\{(z_A,\tz_A) \}_e=\{(N,\tN,X^0,\tX^0,\varphi,\tl\varphi)\}_e$ corresponds to relaxing the uniqueness of the area, and assigning to each face \emph{two areas} $X^0_e$ and $\tX^0_e$ (and their conjugate variables $\varphi_e$ and $\tl\varphi_e$), one for each polyhedral frame. The constraints $H_e$ impose the matching of these areas (as well as reducing $\varphi_e$ and $\tl\varphi_e$ to a single $\xi_e$).

This is the geometric meaning of the constraints $H_e=0$. What we have shown is that the phase space of loop quantum gravity on a fixed graph can be obtained starting from a geometric intepretation of twistors and imposing an \emph{area matching} condition equivalent to say that the twistors are null.

\section{Conclusions}

Let us summarize.
We unraveled a relation between the space $P_*$ of twisted geometries, isomorphic to $T^*\SU(2)$, and null twistors in $\C^4$. Since the phase space of loop quantum gravity on a fixed graph is just the Cartesian product $\bigTimes_e T^*\SU(2)$, our results imply that it can be derived starting from the larger space
 $\bigTimes_e \C^4$, and then imposing the area matching constraint \Ref{H} at each edge.
The derivation can be done in both the usual holonomy-flux parametrization $(g_e,X_e)$ (Proposition 2), or in the twisted geometries parametrization $(N_e, \tN_e, j_e,\xi_e)$ (Proposition 1).

An interesting aspect of the twistor description is that it admits a complete factorization over the \emph{vertices}, as opposed to the edges:
\be\label{vertexfact}
\BigTimes_e \C^4 = \BigTimes_v \C^{2E(v)},
\ee
where $E(v)$ is the valency of the vertex $v$. This result follows straighforwardly once we use the orientation of the edges to uniquely assign $\ket{\bf z}$ to say the source vertex, and $\ket{\bf \tz}$ to the target one. The factorization over the vertices is an interesting spin-off of the twistor description, and can lead to useful applications (e.g. \cite{UN}).

Twistors and twisted geometries form natural spaces that can be associated to a graph. They admit simple geometric interpretations, and are related to loop gravity. Specifically, to the  kinematical (i.e. prior to imposing the Gauss law implementing gauge-invariance) phase space of loop gravity on a fixed graph. For completeness, let us also recall \cite{noi} that gauge-invariance is implemented reducing the space of twisted geometries by the closure conditions
\be\label{C}
C_v \equiv \sum_{e\in v}j_e N_e = 0
\ee 
at each vertex. The resulting space of \emph{closed} twisted geometries is isomorphic to the gauge-invariant phase space of loop gravity, $\times_e T^*SU(2) /\!/ SU(2)^V$.
The variables parametrize it as $\bigTimes_e T^*S^1_e \bigTimes_v S_{\vj_{v}} $, where $T^*S^1$ is the cotangent bundle of a circle, and $S_{\vj_{v}}$ is the space of shapes of a polyhedron, introduced in \cite{Kapovich} and studied in relation to loop gravity in \cite{CF3,FKL}.  
Closed twisted geometries define a local flat metric on each polyhedron. However, this metric is discontinuous: although each face has a unique area, it acquires a different shape when determined from the variables associated to the two polyhedra sharing it, since
there is nothing enforcing a consistent matching of the faces. This discontinuity can be traced back to the fact that the normals carry both intrinsic and extrinsic geometry. 

Finally, for graphs dual to triangulations, the space of closed twisted geometries can be related to the phase space of Regge calculus when one further imposes the gluing or shape matching conditions \cite{BS}. For more discussions on the relation between loop gravity/twisted geometries and discrete gravity, see discussions in \cite{noi,IoCarlo,BiancaJimmy}.

The various phase spaces that can be associated to a graph, and their relations, are summarized by the following scheme:

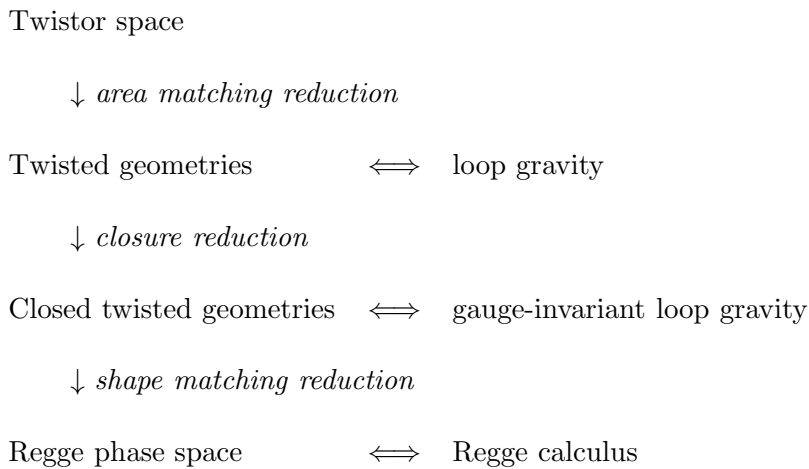
\begin{center}{
\begin{tabular}{lll}
Twistor space & & \\ 
& & \\
\multicolumn{3}{l}{\hspace{.7cm} $\downarrow$ \emph{area matching reduction}} \\
& & \\
Twisted geometries & $\Longleftrightarrow$ & loop gravity  \\
& & \\
\hspace{.7cm} $\downarrow$ \emph{closure reduction} & \\
& & \\
Closed twisted geometries  & 
$\Longleftrightarrow$ & gauge-invariant loop gravity \\
& & \\
\multicolumn{3}{l}{\hspace{.7cm} $\downarrow$ \emph{shape matching reduction}} \\
& & \\
Regge phase space & $\Longleftrightarrow$ & Regge calculus 

\end{tabular}
}\end{center}

\noindent This scheme shows how twisted geometries fit into a larger hierarchy. From top to bottom, we move from larger and simpler spaces, with less intuitive geometrical meaning, to smaller and more constrained spaces, with clearer geometrical meaning. 
The results establish a path between twistors and Regge geometries, via loop gravity.\footnote{For a different relation between twistors and (two-dimensional) Regge calculus, see \cite{Carfora}.} Furthermore, notice also that each phase space but the twistor one is related to a well-known representation of general relativity on a given graph, be it loop gravity or Regge calculus. This raises the intriguing question of whether such a representation can be given directly in terms of twistors. The possibility of defining a  ``twistor gravity'' is a fascinating new direction opened by this new way of looking at loop quantum gravity.

\appendix\section*{Appendix}

In this Appendix we give a direct derivation of \Ref{gzz} that uses explicitly the symplectomorphism \Ref{TSU2P}. To that end, let us briefly review it, refering the reader to \cite{noi} for details. 
We first write the unit vectors as $N = n\tau_3 n^{-1}$, $\tN = \tn\tau_3 \tn^{-1}$, where $n\equiv n(z)\in \SU(2)$ is the Hopf section corresponding to the projection $S^3\mapsto S^2: z\equiv -\bar{z}_1/\bar{z}_0$,
\be
n(z)\equiv \frac1{\sqrt{1+|z|^{2}}}\left(\begin{array}{cc} 1& z \\ -\bar{z} & 1 \end{array}\right).
\ee
Then the (2-to-1) isomorphism is given by 
\begin{subequations}\label{map}\eqa\label{mapX}
(N, \tN, j, \xi) \rightarrow (X, g) \ : \qquad
X &=& j N \\ \label{mapg} 
g &=& n e^{\xi \tau_3} \eps \tl n^{-1}
\neqa\end{subequations}
where $\eps = i\sigma_2$ is the metric tensor in spinor space.\footnote{This was absent in \cite{noi} because the spheres had opposite orientations.} The form of $g$ guarantees that $\tX \equiv j\tN = - g^{-1} X g$. 

The Hopf section defines two families of SU(2) coherent states in the fundamental representation, 
$\ket{n} = n(z)\ket{-}$ and $|n] = -n(z)\ket{+}$, and allows us to bridge between these and  spinors  through the map $z=-\bar{z}_1/\bar{z}_0$,
\bea
\ket{\bf z} &=& \left(\begin{array}{c} z_{0} \\ z_{1} \end{array}\right) = z_0 \left(\begin{array}{c} 1 \\ -\bar{z} \end{array}\right) = \sqrt{\bra{\bf z}{\bf z}\ra} e^{i\arg(z_0)} \, n(z)\ket{+},\\
|{\bf z}] &=&  \left(\begin{array}{c} -\bar{z}_{1} \\ \bar{z}_{0} \end{array}\right) = \bar{z}_{0} \left(\begin{array}{c} z \\ 1 \end{array}\right) =  \sqrt{[{\bf z}|{\bf z}]} e^{-i\arg z_0} \, n(z)\ket{-}.\\
\eea
From these two expressions one immediately finds
\be
g(z_A,\tz_A) = n  e^{\xi \tau_3} \eps \tl n^{-1}
 =n \left(   e^{-\f{i}2\xi} \ket{+} \bra{-}  - e^{\f{i}2\xi}\ket{-} \bra{+} \right) \tn^{-1}
 = \frac{\ket{\bf z}[{\bf \tz}| - |{\bf z}]\bra{\bf \tz}}{ \sqrt{\la {\bf z}|{\bf z} \ra\la {\bf \tz}|{\bf \tz} \ra}},
\ee
where we used $\xi \equiv \xi_{0}= -2(\arg(z_{0}) + \arg( \tz_{0}) )$, and $\epsilon =i\sigma_{2}= \ket{+} \bra{-}  - \ket{-} \bra{+}$.


\end{document}